\documentclass[10pt, final, onecolumn, twoside,romanappendices]{IEEEtran}
\usepackage{cite}
\usepackage[cmex10]{amsmath}
\usepackage{amsthm}
\usepackage{amssymb}
\usepackage{winsnotation}
\usepackage{acronym}  % make an acronym
\usepackage{psfrag}
\usepackage{color}  % make colors
\usepackage[usenames,dvipsnames]{xcolor}
\usepackage{algorithm}
\usepackage{algpseudocode}
\usepackage[utf8]{inputenc}
\usepackage[english]{babel}
\usepackage{physics}
\usepackage{pdfsync}
\usepackage[percent]{overpic}
\usepackage{tikz}
\usepackage{pgfplots}
\usepackage{pstricks}
\usepackage[free-standing-units]{siunitx}
\usepackage{circuitikz}
\usepackage{float}
\usepackage{schemabloc}
\usepackage{mathtools}
\usepackage{xfrac}
\usepackage{subfig}
\usepackage{multirow}
\usepackage{accents}
\usepackage[font=small]{caption}
\usepackage{dblfloatfix}    % To enable figures at the bottom of page
\usepackage{pict2e}
\usepackage{listings}
\usepackage{setspace}
\usetikzlibrary{matrix}
\usetikzlibrary{circuits}
\usetikzlibrary{arrows,positioning}

\newtheorem{innercustomgeneric}{\customgenericname}
\providecommand{\customgenericname}{}
\newcommand{\newcustomtheorem}[2]{%
  \newenvironment{#1}[1]
  {%
   \renewcommand\customgenericname{#2}%
   \renewcommand\theinnercustomgeneric{##1}%
   \innercustomgeneric
  }
  {\endinnercustomgeneric}
}

\newcustomtheorem{customthm}{Theorem}
\newcustomtheorem{customlemma}{Lemma}

\definecolor{0}{HTML}{54FF00}
\definecolor{1}{HTML}{FFFFFF}
\definecolor{2}{HTML}{FF0000}
\definecolor{3}{HTML}{0048FF}
\definecolor{4}{HTML}{EA00FF}
\definecolor{5}{HTML}{DC00F0}
\definecolor{6}{HTML}{FFFF00}
\definecolor{7}{HTML}{1E90FF}
\definecolor{8}{HTML}{FF1493}
\definecolor{9}{HTML}{00FFFF}
\definecolor{10}{HTML}{C0C0C0}

\usetikzlibrary{shapes,arrows}
\usetikzlibrary{shapes.misc}
\tikzstyle{line} = [draw, -latex']

\tikzstyle{block} = [draw, fill=white, rectangle,
    minimum height=4.5em, minimum width=6em]
\tikzstyle{sum} = [draw, fill=white, circle, node distance=1cm]
\tikzstyle{input} = [coordinate]
\tikzstyle{output} = [coordinate]
\tikzstyle{pinstyle} = [pin edge={to-,thin,black}]

\tikzset{%
  wireless/.pic={
      \draw [->] (0,0) -| (.5,#1);
    \foreach \r in {.1,.2,.3}
      \draw (.6,#1) ++ (60:\r) arc (60:-60:\r);
  },
  vdots/.pic={
    \foreach \i in {-.1,0,.1}
      \fill (.25,\i) circle [radius=.75pt];
  },
  block/.style={
    shape=rectangle,
    minimum width=6em,
    minimum height=4.5em,
    draw
  },
  RF chain/.style 2 args={
    block,
    node contents=RF chain,
    append after command={
      \pgfextra{\pgfnodealias{@}{\tikzlastnode}}
      (@.north #1) [yshift=-.25cm] pic [#2] {wireless=.5}
    }
  },
  MIMO RF chain east/.style={RF chain={east}{xscale=1}},
  MIMO RF chain west/.style={RF chain={west}{xscale=-1}},
}

\tikzstyle{multiply}=[draw,circle,minimum width=0.5 cm]

\makeatletter
\def\BState{\State\hskip-\ALG@thistlm}
\makeatother

\DeclareMathAlphabet{\pazocal}{OMS}{zplm}{m}{n}

\newcommand{\bd}{\begin{description}}
\newcommand{\ed}{\end{description}}
\newcommand{\be}{\begin{enumerate}}
\newcommand{\ee}{\end{enumerate}}
\newcommand{\bi}{\begin{itemize}}
\newcommand{\ei}{\end{itemize}}
\newcommand{\bl}{\begin{list}}
\newcommand{\el}{\end{list}}
\newcommand{\bt}{\begin{tabbing}}
\newcommand{\et}{\end{tabbing}}

\definecolor{BLUE}{rgb}{0,0,1}

\usepackage{acronym}  % make an acronym
\acrodef{mud}[MUD]{multi-user detection}
\acrodef{sud}[SUD]{single-user detection}
\acrodef{mu}[MU]{multiuser}
\acrodef{bs}[BS]{base station}
\acrodef{ma}[MA]{multiple access}
\acrodef{iot}[IoT]{Internet-of-Things}
\acrodef{iov}[IoV]{Internet of Vehicles}
\acrodef{ra}[RA]{random access}
\acrodef{sb}[SB]{signature-based}
\acrodef{urllc}[uRLLC]{Ultra Reliable and Low Latency Communications}
\acrodef{ap}[BS]{base station}
\acrodef{rach}[RACH]{RA channel}
\acrodef{phy}[PHY]{physical}
\acrodef{noma}[NOMA]{non-orthogonal multiple access}
\acrodef{sbra}[SBRA]{signature-based random access}
\acrodef{nora}[NORA]{non-orthogonal random access}
\acrodef{toa}[ToA]{time-of-arrival}
\acrodef{sic}[SIC]{successive interference cancellation}
\acrodef{rar}[RAR]{random access response}
\acrodef{sar}[SAR]{scheduled access request}
\acrodef{rsma}[RSMA]{resource spread multiple access}
\acrodef{rdma}[RDMA]{repetition division multiple access}
\acrodef{goca}[GOCA]{group orthogonal coded access}
\acrodef{igma}[IGMA]{interleave-grid multiple access}
\acrodef{sdma}[SDMA]{signature-division multiple access}
\acrodef{gf}[GF]{grant-free}
\acrodef{rg}[RG]{request-grant}
\acrodef{mui}[MUI]{multiuser interference}
\acrodef{pdma}[PDMA]{pattern division multiple access}
\acrodef{scma}[SCMA]{sparse code multiple access}
\acrodef{lds}[LDS]{low-density spreading}
\acrodef{nlds}[NLDS]{non-low-density spreading}
\acrodef{musa}[MUSA]{multi-user shared access}
\acrodef{rpma}[RPMA]{random phase multiple access}
\acrodef{mpa}[MPA]{message passing algorithm}
\acrodef{map}[MAP]{maximum a posteriori probability}
\acrodef{pic}[PIC]{parallel interference cancellation}
\acrodef{mmse}[MMSE]{minimum mean square error}
\acrodef{mf}[MF]{matched filter}
\acrodef{ic}[IC]{interference cancellation}
\acrodef{bler}[BER]{bit error rate}
\acrodef{sinr}[SINR]{signal-to-interference-plus-noise ratio}
\acrodef{ese}[ESE]{elementary signal estimator}
\acrodef{idma}[IDMA]{interleave-division multiple access}
\acrodef{fec}[FEC]{forward error correction}
\acrodef{np}[NP]{nondeterministic polynomial time}
\acrodef{llr}[LLR]{log-likelihood-ratio}
\acrodef{cdma}[CDMA]{code division multiple access}
\acrodef{noca}[NOCA]{non-orthogonal coded access}
\acrodef{mc}[MC]{multicarrier}
\acrodef{ds}[DS]{direct sequence}
\acrodef{qpsk}[QPSK]{quadrature phase-shift keying}
\acrodef{lte}[LTE]{Long-Term Evolution}
\acrodef{cp}[CP]{cyclic prefix}
\acrodef{ncma}[NCMA]{non-orthogonal coded multiple access}
\acrodef{ici}[ICI]{inter channel interference}
\acrodef{sc}[SC]{single-carrier}
\acrodef{papr}[PAPR]{peak-to-average power ratio}
\acrodef{mac}[MAC]{medium access control}
\acrodef{csi}[CSI]{channel state information}
\acrodef{crc}[CRC]{cyclic redundancy check}
\acrodef{cs}[CS]{compressive sensing}
\acrodef{s-noma}[S-NOMA]{signature-based NOMA}
\acrodef{mtc}[MTC]{machine-type communications}
\acrodef{oma}[OMA]{orthogonal multiple access}
\acrodef{epa}[EPA]{expectation propagation algorithm}
\acrodef{dof}[DoF]{degrees of freedom}
\acrodef{sf}[SF]{spreading factor}
\acrodef{ovf}[OLF]{overloading factor}
\acrodef{pn}[PN]{pseudo-noise}
\acrodef{lpwa}[LPWA]{low power wide area}
\acrodef{fn}[FN]{function node}
\acrodef{vn}[VN]{variable node}
\acrodef{mmtc}[mMTC]{massive machine-type communications}

\newcommand{\paperTitle}{Signature-based Non-orthogonal Multiple Access (S-NOMA) for Massive Machine-Type Communications in 5G}

\begin{document}
\renewcommand{\figurename}{Fig.}

%%%%% UNCOMMENT THIS LINE IF YOU HAVE 2-column version %%%%%%
%\twocolumn
%%%%% UNCOMMENT THIS LINE IF YOU HAVE 2-column version %%%%%%

%---------------
% Release notes:
%   Jan 25, 2012 -- Ae added an example of how to use acronym package
%   Jan  5, 2012 -- Ae updated the statement above the sponsor (in title),
%                   made the cover sheet 1-column (in section cover sheet), and
%                   added a command "\paperTitle" (for section cover sheet and title).
%   June 2, 2011 -- Ae updated the questions in "WGroup Research Paper Catechism"
%                   (using the questions from Prof. Win and Andrea Conti)
%---------------

%---------------------------------------------------------------------------%
%                     title, title footnote, header                         %
%---------------------------------------------------------------------------%

\title{\paperTitle}

% Uncomment this line, if it's an invited paper
% \IEEEspecialpapernotice{(Invited Paper)}

% author names, IEEE memberships, corresponding address, title footnote %

\author{
%%%%%%%% uncomment this section for a 2-column formt %%%%%%%
%%%%%%%% [begin] %%%%%%%%
	%\vspace{0.2cm}
%%%%%%%% [end] %%%%%%%%	
Mostafa Mohammadkarimi,~\IEEEmembership{Member,~IEEE,}
        Muhammad Ahmad Raza,~\IEEEmembership{Student Member,~IEEE,}
        \\
        Octavia A. Dobre, ~\IEEEmembership{Senior Member,~IEEE}
        \thanks{M. Mohammadkarimi is with the Department of Electrical and Computer Engineering, University of Alberta, Edmonton, AB, Canada.
(e-mail: mostafa.mohammadkarimi@ualberta.ca). M. A. Raza, and O. A. Dobre are with the Department of Electrical and Computer Engineering, Memorial University, St. John's, NL, Canada (e-mail: odobre@mun.ca and maraza@mun.ca).

This work was supported by the Natural Sciences and Engineering Research Council of Canada (NSERC), through its Discovery program.}}
\maketitle
\acresetall	
\null
%---------------------------------------------------------------------------%
%                           abstract and key words                          %
%---------------------------------------------------------------------------%
\doublespacing
\begin{abstract}
 The problem of providing massive connectivity in \ac{iot} with a limited number of available resources motivates the \ac{noma} solutions.
  In this article, we provide a comprehensive review of the \ac{s-noma} schemes as potential candidates for IoT. The signature in \ac{s-noma} represents the way the data stream of an active device is spread over available resources in a non-orthogonal manner. It can be designed based on device-specific codebook structures, delay patterns, spreading sequences, interleaving patterns, and scrambling sequences.
 Additionally,
we present the detection algorithms employed to decode each device's data from non-orthogonally superimposed signals at the receiver.
 The bit error rate of different \ac{s-noma} schemes is simulated in impulsive noise environments, which can be important in machine-type communications. Simulation results show that the performance of the \ac{s-noma} schemes degrades under such conditions.
 Finally, research challenges in \ac{s-noma} oriented \ac{iot} are presented.
\end{abstract}
\acresetall		
%---------------------------------------------------------------------------%
%                                Introduction                               %
%---------------------------------------------------------------------------%
\section{Introduction}\label{sec:intro}
With the explosive demand for higher data rate, last decades have witnessed the evolution of the mobile wireless communication systems from the first generation to the fourth generation (4G). However, the objective of the future wireless communication systems (5G and beyond) will not only be to provide services with higher data rate. One of the main goals of these systems is to support \ac{mmtc} within the paradigm of the \ac{iot}.
In \ac{mmtc} a large  number  of  devices  are  connected  to
one \ac{ap}, each  transmitting  sporadically  a  small
payload information
without any human interactions.
According to a forecast by Ericson, several billions of \ac{iot} devices will be installed by 2020\cite{shirvanimoghaddam2017massive}.

 To support the ubiquitous connectivity, new advancements of communication technologies and protocols are required. One of the fundamental aspects in this regard is the \ac{mac}. Since the number of available resources, such as frequency and time, are limited, conventional \ac{oma} schemes cannot support the massive connectivity in \ac{iot}. The large gap between the available resources and required massive connections motivates the research community to investigate \ac{noma} solutions. In this regard, several schemes have been considered by the 3GPP New Radio standard
%\footnote{Although the 3GPP Release-15 for the 5G New Radio was released in June 2018, according to \cite{3gprsmayu55geterter} (section 5.1), the standardization of \ac{noma} schemes to be used in the uplink is not complete yet.}
 and the \ac{lpwa} technologies
\cite{yang2017uplink,myers2010random,chandra2018unveiling}.

\begin{figure*}
  \centering
  \includegraphics[width=18cm]{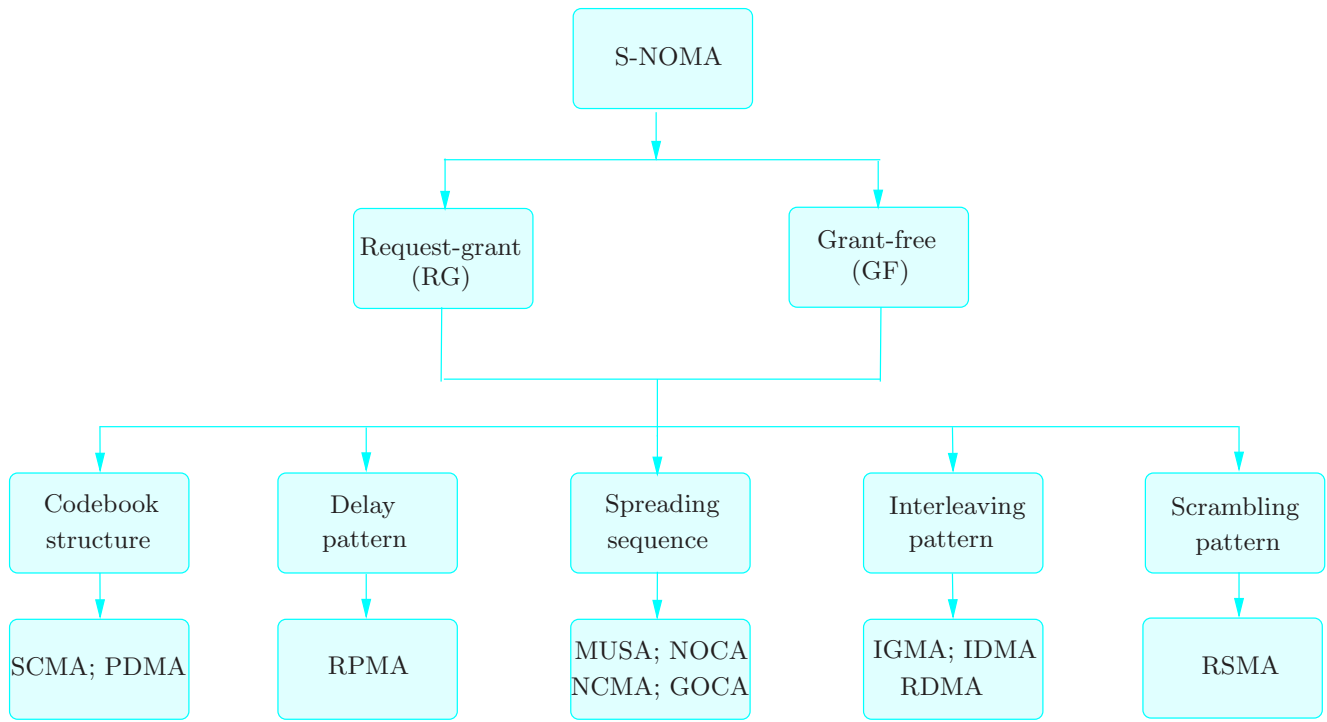}
  \caption{Taxonomy of \ac{s-noma} schemes. SCMA: sparse code multiple access; PDMA: pattern division multiple access;
   RPMA: random phase multiple access, MUSA: multi-user shared access, NOCA: non-orthogonal coded access, NCMA: non-orthogonal coded
multiple access, GOCA: group orthogonal coded access; IGMA: interleave-grid
multiple access, IDMA: interleave-division multiple-access, RDMA: repetition division multiple access; RSMA: resource spread
multiple access.}\label{yyyiioeoea}
\end{figure*}

The \ac{s-noma} represents a class of schemes in which the data from each active device is spread over available resources in a non-orthogonal manner by using a set of predefined signatures. This
provides
additional \ac{dof} for device separation, which leads to a significantly high \ac{ovf}.\footnote{The \ac{ovf} is defined as the ratio of the number of overloaded signals to the number of orthogonal resources.}
The signatures can be designed based on device-specific codebook structures, delay patterns, spreading sequences, interleaving patterns, and scrambling sequences.
At the \ac{ap}, advanced activity detection and \ac{mud} algorithms are employed to decode each device's data from the non-orthogonally superimposed signals. The joint design of signatures and \ac{mud} receiver is performed to improve the resilience of each device's signal against \ac{mui} and to obtain a low-complexity \ac{mud} receiver \cite{5gsignal}.
Fig. \ref{yyyiioeoea} presents a taxonomy of the \ac{s-noma} schemes. In addition to the classification based on signature, each scheme can be implemented following either a \ac{gf} or \ac{rg} approach.

This article provides a comprehensive review of different \ac{s-noma} schemes, which can support massive uplink connectivity in \ac{iot}. Both \ac{gf} and  \ac{rg} \ac{s-noma} scenarios are discussed. Moreover, the \ac{bler} of various \ac{s-noma} schemes is investigated through simulation in the presence of impulsive noise. Finally,  potential research challenges for \ac{s-noma} in \ac{iot} are presented.
\section{\ac{s-noma}: Random Access (RA) Procedure}
The conventional \ac{rg}- and \ac{gf}-based uplink multiple access schemes cannot support massive connectivity in machine-type communications since
the maximum number of
supported devices is limited by the number
and scheduling granularity of orthogonal
resources.
A promising way to scale up the uplink connectivity in \ac{iot} applications is to use S-NOMA in both RG- and GF-based MAC schemes.
A brief presentation of the conventional \ac{ra} procedure, as well as of the \ac{rg} and \ac{gf} \ac{s-noma}, is provided in the following.
\subsection{Collision in the Conventional \ac{ra} Procedure}
Fig. 2a shows the four-step handshake \ac{ra} procedure in the conventional \ac{rg}-based multiple access employed in \ac{lte}.
In the RA procedure, during the \ac{rach} stage, more than one device can select the same orthogonal preambles. Identical preamble selection leads to preamble collision and makes this RG-based multiple access scheme ineffective for massive uplink connectivity.
%In this procedure,
%more than one device can select the same orthogonal
%preamble in the \ac{rach} process, which leads to preamble collision and inadequacy for massive uplink connectivity.
Similarly, the conventional \ac{gf}-based multiple access schemes employed in wireless networks, such as ALOHA, slotted ALOHA, and their variants suffer from low throughput due to the high probability of collision
in massive connectivity.
% In case of collision, the devices
%repeat the preamble transmission in the next available \ac{ra}
%slot in the \ac{rg}-based schemes, and retransmit their packet after a random back-off time in the \ac{gf}-based schemes. Frequent preamble and packet  retransmissions lead
%to network congestion, increasing delays, high
%energy consumption, excessive signaling overhead, and radio
%resource wastage.
Furthermore, in both \ac{rg}- and \ac{gf}-based schemes, successful data transmission is achieved through \ac{oma}; however,
there is a limited number of orthogonal resources to be allocated
to a very large number of devices
 %\cite{chandra2018unveiling}.
\subsection{Request-grant \ac{s-noma}}
The main idea behind \ac{rg} \ac{s-noma} is to reduce the probability of collision and increase the \ac{ovf} in the \ac{rach} process of the \ac{ra} procedure
by employing advanced \ac{mud}.
Recently, an \ac{rg} \ac{s-noma} scheme was proposed in \cite{liang2017non}, where the \ac{ap} utilizes the difference in the \ac{toa} to
identify multiple devices with identical preambles, as shown in Fig. 2b.
This represents a modification of the \ac{lte} \ac{ra} procedure, by including a delay pattern \ac{s-noma} scheme in the \ac{rach} process.
 As seen, after preamble transmission,
the \ac{ap} implements multi-preamble detection to detect the delayed versions of the transmitted preambles.
Then, it considers the devices with
detected collisions, i.e., devices with identical preamble, as a \ac{nora} group and responds to the group
with a \ac{rar} message.
Based on
the channel conditions obtained through preamble detection, the \ac{rar} message includes the power back-off values for devices in the \ac{nora} group.
Each device estimates its distance from the \ac{ap} based
on the received signal strength of cell-specific reference
signal to obtain its \ac{toa}.
Based on the estimated \ac{toa}, each device searches the
closest \ac{toa} value in the \ac{rar} message.
The power back-off information is utilized for the power-domain multiplexing at
the device-side for \ac{sar}, as well as for performing \ac{sic} at the \ac{ap}.
%The decoding order of
%devices in a \ac{nora} group is consistent with the power back-off
%order, i.e., the devices with the strongest received power will be
%decoded first.
%first, which is reasonable due to the SIC \ac{mud}.
\begin{figure*}[]
\centering
\captionsetup[subfloat]{position=below,labelformat=empty}
\subfloat[({\bf{a}}) Conventional \ac{rg} \ac{oma} in the \ac{lte} \cite{liang2017non}.]{
  \includegraphics[width=6.95cm]{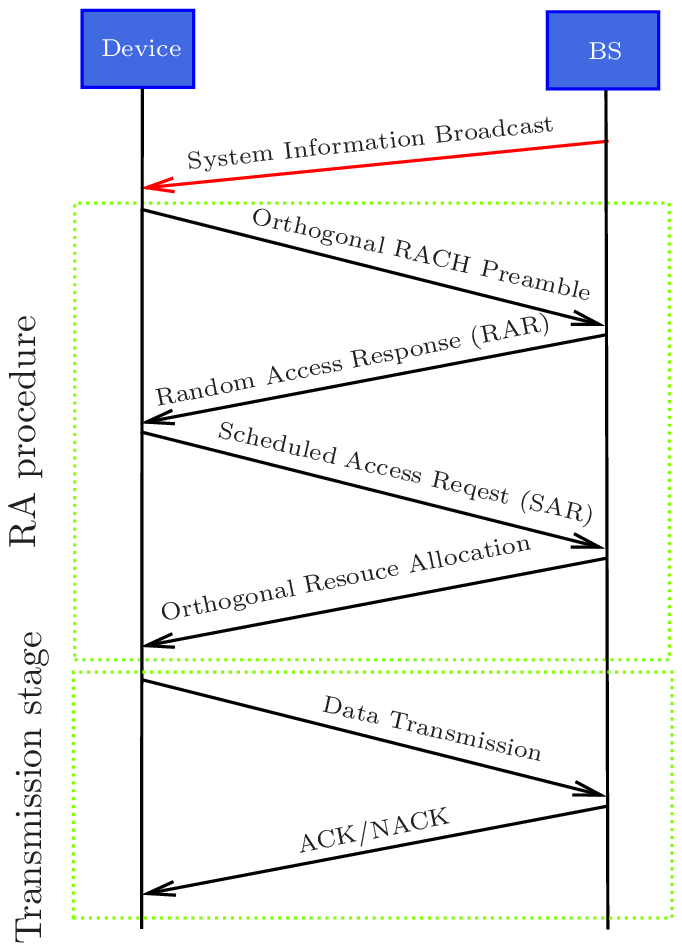}\label{fig:tttyyyyy44444444444}
  \label{fig:evaluation:revenue}%
}\
\subfloat[ ({\bf{c}}) \ac{gf} \ac{s-noma} \cite{3gpmusa}.]{%
  \includegraphics[width=6.95cm]{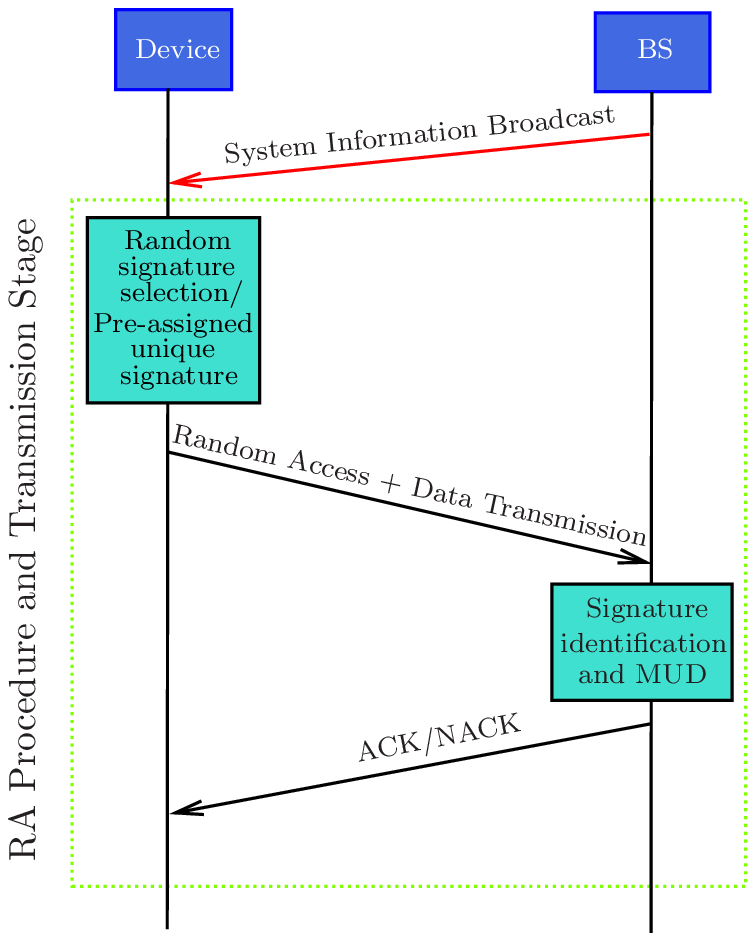}\label{fig:tttyyyyy8859w}
  \label{fig:evaluation:avgPrice}%
}\
\subfloat[\hspace{-1cm}({\bf{b}}) \ac{rg} \ac{s-noma} (A modification of the \ac{lte} channel access) \cite{liang2017non}.]{%
  \includegraphics[width=7.2cm]{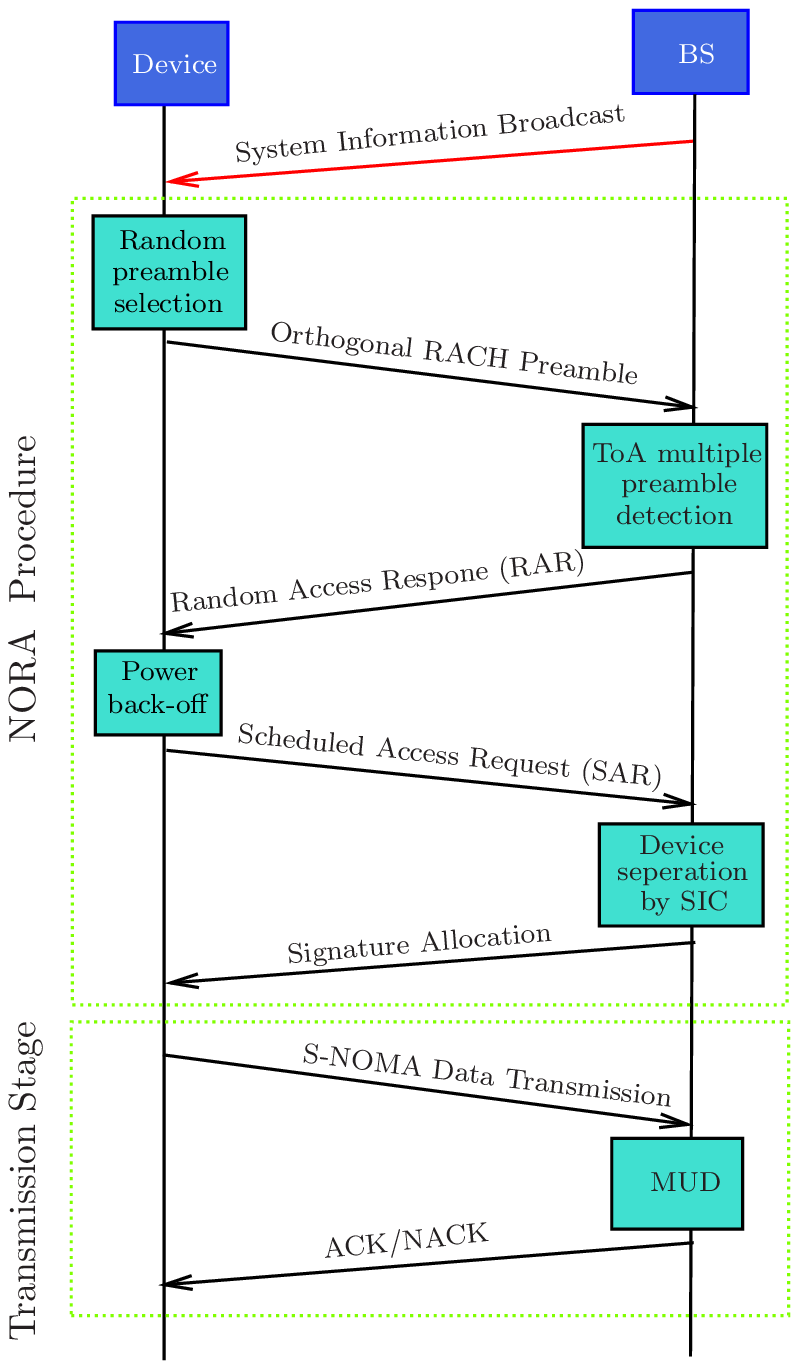}\label{fig:tttyyyyy88e}
%  \label{fig:evaluation:avgPrice}%
}
\caption{\ac{ra} procedure and transmission stage in  a) conventional \ac{rg} \ac{oma} \cite{liang2017non} b) \ac{rg} \ac{s-noma} \cite{liang2017non}  and c) \ac{gf} \ac{s-noma} \cite{3gpmusa}. ACK: acknowledgment; NACK: negative acknowledgment.}\label{fig:mul776501}
\end{figure*}
A more general approach to \ac{nora} grouping is to directly employ the \ac{s-noma} signatures at the RACH stage to further reduce the probability of collision and increase the \ac{ovf}.
\\
Finally, in the transmission stage, the \ac{ap} allocates resources to those devices that successfully transmitted \ac{sar}.
As distinct signatures are allocated to devices in this stage, collisions do not occur at the expense of higher control signaling and higher complexity \ac{mud} in the \ac{ra} procedure when compared with the \ac{gf} \ac{s-noma} scheme; this will be described in the next section.

%It deterministically allocates the signatures to devices, i.e., different devices are not allocated identical signatures. Hence, collisions do not occur in the transmission stage at the expense of higher control signaling and higher complexity \ac{mud} in the \ac{ra} procedure when compared with the \ac{gf} \ac{s-noma} scheme, which will be described in the next section.
\subsection{Grant-free \ac{s-noma}}
In grant-free \ac{s-noma}, each active device directly transmits its packets to the \ac{bs} without waiting for any permission.
Fig. 2c shows the joint \ac{ra} procedure and transmission stage in \ac{gf} \ac{s-noma}. Both random signature selection and pre-assigned signature allocation schemes have been investigated for \ac{gf} \ac{s-noma} \cite{3gpmusa}. In the former scheme,
each device randomly selects a signature
from a large pool for link establishment. On the other hand, under the latter scheme, each device is pre-assigned a unique signature used for all transmission slots. This signature thus serves as the device identification, as well.
These schemes increase the spectral efficiency
and number of served devices through the simultaneous
non-orthogonal \ac{rach} and data transmission at devices,
as well as
joint signature identification, and detection of device activity and data at the \ac{ap}.
%The randomness in signature selection can cause collision when different devices pick the same signature. However, using a signature pool with large cardinality significantly reduces the probability of collision, leading to higher throughput.
%Furthermore, when signatures are available (e.g., for sporadic traffic), a device can utilize multiple signatures either to improve the reliability of communication or increase the data rate.
\ac{gf} \ac{s-noma} allows asynchronous, non-orthogonal
contention-based access that is well
suited for sporadic uplink transmissions of small
data bursts common in \ac{iot} applications.
In order to support a large number of devices, the network can be partitioned into groups, with a signature pool allocated to each group.
\section{Signature Design in \ac{s-noma}}
The \ac{s-noma} \ac{ovf} and complexity of the \ac{mud} receiver at the \ac{ap} depend on the design of signatures.
To design efficient signatures, different operations, such as linear spreading, multi-dimensional modulation, interleaving, and scrambling can be employed \cite{5gsignal}.
In this section, we overview how these operations are employed at bit-level and/or symbol-level to provide efficient signatures.
\subsection{Codebook Structure \ac{s-noma}}
\begin{figure*}
\hspace{5em}
\includegraphics[width=18cm]{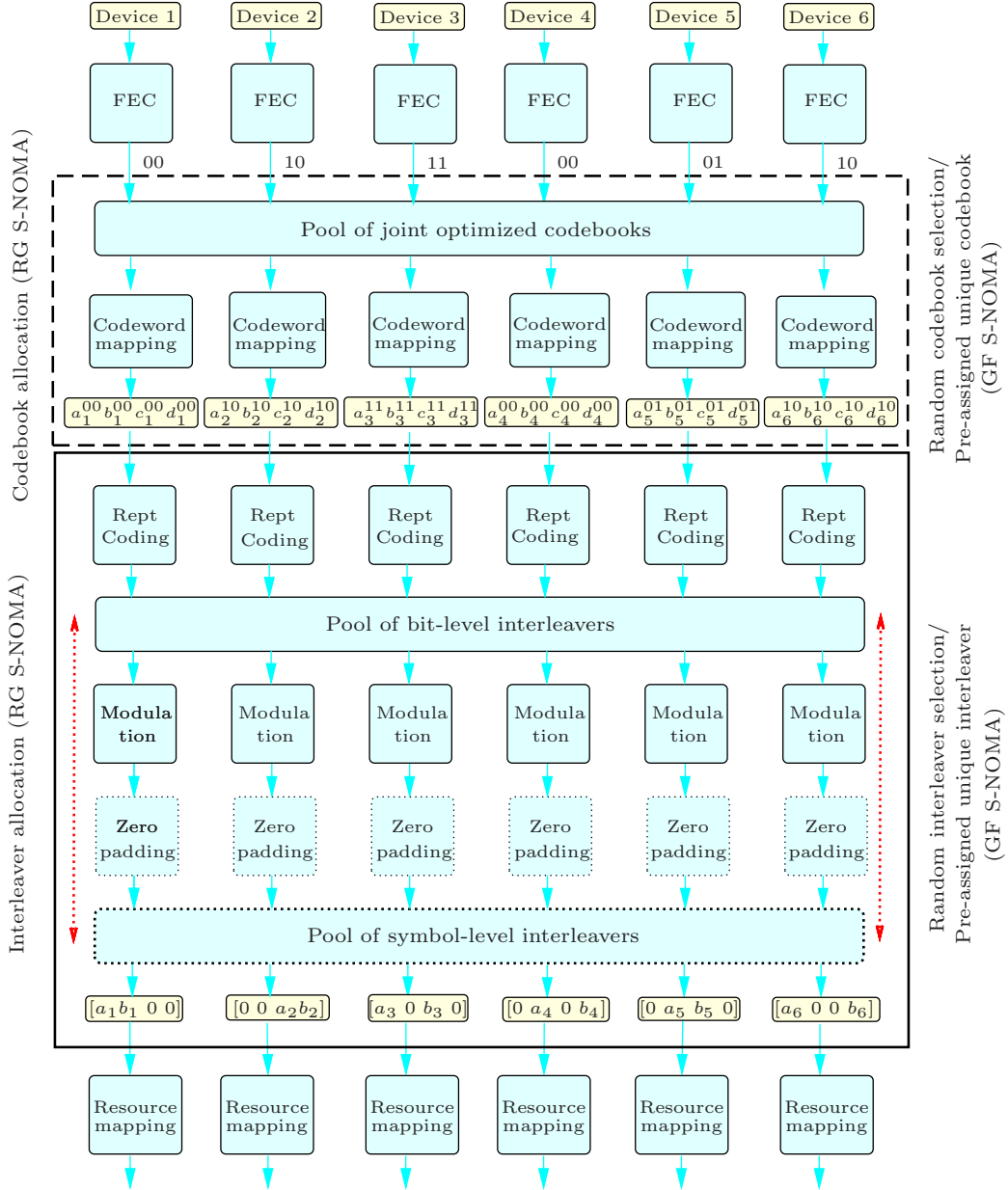}%
\caption{Codebook structure \ac{s-noma} with \ac{sf} 4 and \ac{ovf} $150\%$ (Dashed box). Every two bits are directly mapped into a codeword with length 4 in a codebook.
Interleaving pattern \ac{s-noma}  with \ac{sf} 4 and \ac{ovf} $150\%$ (Solid box). The codewords consist of two zero entries through zero-padding and $a_i$ and $b_i$, which are the modulated symbols of the $i$th device. Rept stands for repetition.}\label{fig:mul776501er}
\end{figure*}

The main idea behind the codebook structure \ac{s-noma} is the direct mapping of each device's data stream
 onto a codeword in a structured codebook.
The sparsity pattern used in codebooks is different in order to facilitate device separation and reduce \ac{mui}, while it is kept the same in codewords belonging to the same codebook.
The sparsity pattern of codebooks defines the signatures, and the number of non-zero elements in a codeword represents the number of orthogonal resources used for transmission; this equals the diversity order.
The \ac{sf} is equivalent to the length
of a codeword, and the \ac{ovf} is determined by
the ratio of the number of multiplexed devices in a group to the spreading
factor.
Two potential schemes in this category are \ac{scma} and \ac{pdma}.
Fig. \ref{fig:mul776501er} (dashed box) illustrates the codebook structure \ac{s-noma} in which every two bits are directly mapped into a codeword with length four.

{\it 1) {\ac{scma}}}
provides the same diversity order for different devices.
Its implementation is similar to LTE, with a key difference of the joint design of modulation and low-density spreading.
This design is implemented through a multi-dimensional modulation with lower number of projection points. A multi-dimensional modulation is
simply the mapping of the input coded bits onto the points in the multiple complex dimensions (containing both I and Q).
By employing multi-dimensional modulation, the \ac{scma} codewords provide both diversity and shaping gain \cite{y4455566777H}.

{\it 2) {\ac{pdma}}}
%is inspired by the ideas of unequal transmission diversity and sparse coding.\ac{pdma}
sparsely maps each devices's data
stream onto a group of resources according
to a \ac{pdma} codebook to realize disparate transmission diversity
order.
A resource group can consist of
time, frequency, and spatial resources, or any combination
of them.  Data
of multiple devices can be multiplexed onto the same resource
group with a different \ac{pdma} pattern.
In general, if $N$ is the size of the orthogonal resource group, there are $2^N-1$ possible binary vectors for the transmission pattern.   Assuming $K$ is the number of devices, we can thus choose $K$ transmission pattern vectors out of the $2^N-1$ candidates to construct codebooks.
The transmission patterns are chosen to maximize the diversity order while minimizing the overlaps among devices.
The same transmission diversity order can be considered in PDMA while minimizing the maximum inner product between pattern vectors.
%In PDMA, the same transmission diversity order can be considered; in this case, the maximum
%inner product between pattern vectors with the same diversity order should be minimized.
After transmission pattern selection, all codewords in a codebook are designed following the same transmission pattern vector \cite{chen2017pattern}.
%It is obvious that the selection of the transmission pattern impacts the performance and complexity of the \ac{mud} receiver \cite{chen2017pattern}.

\subsection{Interleaving Pattern \ac{s-noma}}
The key characteristic of interleaving \ac{s-noma}
is that different interleavers along with repetition and/or low-rate \ac{fec} coding are employed for device separation.
The length of the repetition code and/or the inverse of code rate determines the \ac{sf}, while the interleaver patterns define the signatures. Superposition transmission with different interleaving patterns results in
interference averaging, and thus, lower \ac{mui}. Three schemes in this category are: \ac{igma}, \ac{idma}, and \ac{rdma}.
Fig. \ref{fig:mul776501er} (solid box) illustrates the interleaving pattern \ac{s-noma} concept.

{\it 1) {\ac{igma}}}
provides high \ac{dof} for device separation by
employing different bit-level interleavers, different symbol-level grid mapping patterns, and a combination of bit-level interleavers and grid mapping patterns.
%The interleavers and/or grid mapping patterns can be selected randomly (\ac{gf}-\ac{igma}) or deterministically (\ac{rg}-\ac{igma}).
%The high \ac{dof} result in a large number of distinct signatures to support massive connectivity.
In \ac{igma}, low code-rate \ac{fec} coding and repetition coding can be considered as a spreading technique.
%; thus, it leads to higher \ac{sinr} in multiuser transmission.
After
channel coding and repetition, the bit-level interleaver is employed to make
the transmission bits randomly distributed.
Then,
in the grid mapping process, sparse mapping based on zero padding and symbol-level interleaving are introduced to provide another dimension for device multiplexing. Sparse mapping leads to further \ac{mui} reduction,
and the symbol-level interleaving randomizes the symbol sequence order, which may further bring benefits in terms of combating frequency-selective fading and inter-channel interference. \ac{igma} can achieve
coding and diversity
gains \cite{yang2017uplink}.

{\it 2) {\ac{idma}}}
is a special case of \ac{igma}, where
different devices are
distinguished by bit-level interleavers. The bit-level interleaver, along with repetition and/or very low-rate \ac{fec} coding, enables dispersion of data
across a long bit stream, which can provide time-frequency diversity.
The main advantages of \ac{idma} consist of high \ac{ovf}
and low \ac{bler} when it uses very low-rate \ac{fec}. Also,
the interleaved low-rate codes with a simple chip-by-chip
iterative decoding strategy could achieve the capacity of a
Gaussian multiple access channel \cite{ping2006interleave}.

{{\it 3) {\ac{rdma}}}
employs different simple cyclic-shift repetition patterns in the symbol-level to design device-specific signatures. The cyclic-shift repetition behaves as a randomizer (interleaver), and
provides both time and frequency diversity. Moreover, the
repetition patterns lead to randomized \ac{mui} in both time and frequency domain for each repeated modulated symbol. On the contrary to
\ac{idma}, random interleaver is not used in \ac{rdma} \cite{3gp4}.
%The receiver employs \ac{sic}, which provides a good trade-off between complexity and performance \cite{3gp4}.

\begin{figure*}
\hspace{5em}
\includegraphics[width=17cm]{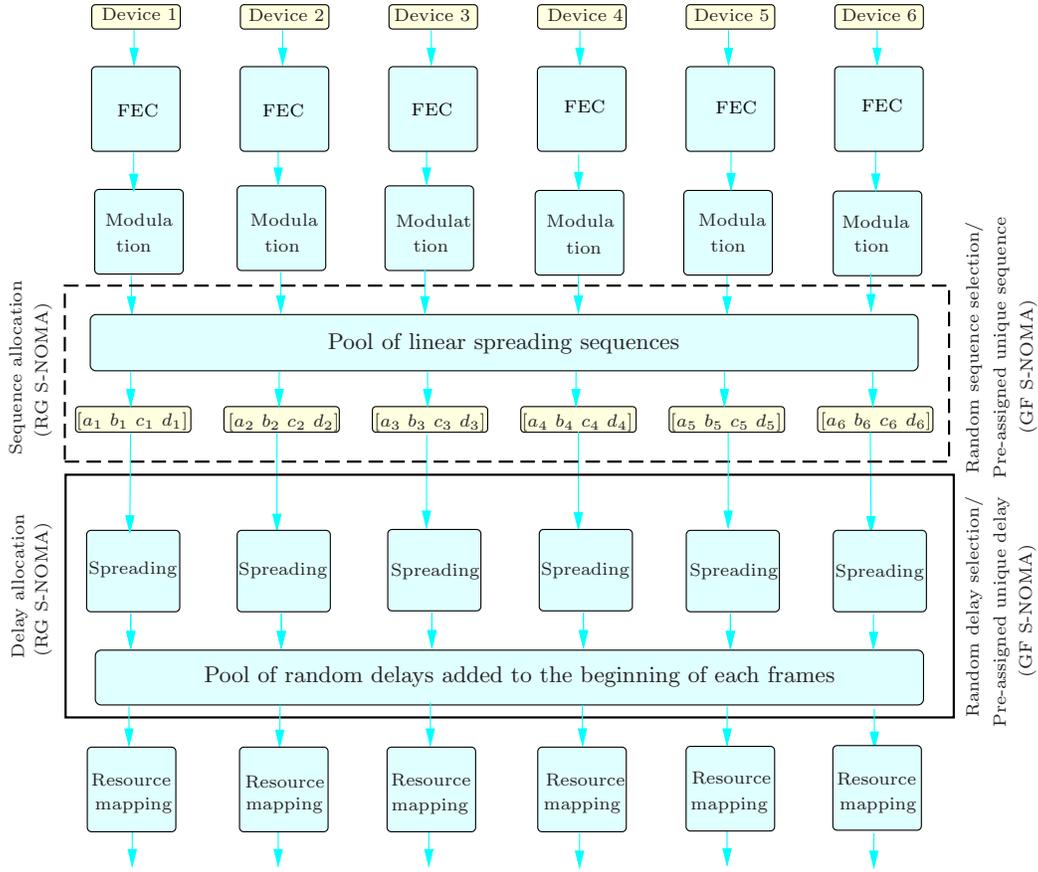}%
\caption{Spreading sequence \ac{s-noma} with \ac{sf} 4 and \ac{ovf} $150\%$ (Dashed box). The entries of the $i$th codeword, i.e., $[a_i \ b_i \ c_i \ d_i]$, represent the product of a modulated symbol and the spreading sequence of the $i$th device. Delay pattern \ac{s-noma} (Solid box); devices employ identical spreading sequences, such as the Gold code. }\label{fig:mul776501yty}
\end{figure*}

\subsection{Spreading Sequence \ac{s-noma}}
The idea behind
spreading sequence \ac{s-noma} is that low cross-correlation real/complex-valued spreading sequences are employed as device-specific signatures in order to enable non-orthogonal transmission\footnote{Complex sequences provide additional \ac{dof} for device separation.}.
%Linear spreading is the simple weighted repetition of the modulated symbols, obtained by multiplying the modulated symbol and spreading sequence.
Since the spreading sequences are not necessarily orthogonal, this leads to high \ac{dof} for device separation \cite{5gsignal}.
%When compared with \ac{scma},
%a linear spreading technique is employed instead of multi-dimensional
%modulation.
%These real/complex-valued sequences are generated by linear spreading i.e., by the weighted repetition of modulated symbols.
After linear spreading, symbols are directly mapped to orthogonal resources.
%In this category, a family of short length real/complex-valued sequences with low cross-correlation is usually chosen
%to enable a simple \ac{mui} cancellation.
%It is obvious that complex sequences provide additional \ac{dof} for device separation.
%Depending on whether or not there exist zero entries in the
%spreading sequence,
While both \ac{lds}- and non-\ac{lds} sequences can be considered, the former is preferable since it can efficiently  mitigate \ac{mui}.
The \ac{ovf} is defined the ratio of the number of devices to the length of the
spreading sequence.
Existing \ac{s-noma} schemes in this category are:
\ac{musa}, \ac{noca}, \ac{ncma}, and \ac{goca}.
The main difference between these schemes is related to the spreading sequences used for device separation.
In Fig. \ref{fig:mul776501yty} (dashed box), the concept of the spreading sequence \ac{s-noma} is illustrated.

\subsubsection{\ac{musa}}
Each device's modulated
data symbols are spread by a special family of low cross-correlation complex-valued spreading sequences,
which can facilitate \ac{mud} based on \ac{sic} implementation. The real and imaginary parts of the employed non-binary spreading sequence are
drawn from an $M$-level value set ($M\geq 2$) with uniform distribution. As an example,
 $\{-1,0,1\}$ is a popular set for \ac{musa}, leading to $9^L$ sequences, with $L$ as the sequence length.
The main advantage of \ac{musa} consists of its high \ac{ovf}; this can be as high as $700\%$ in multipath fading channel.
Moreover, the blind detection based on the \ac{sic} receiver makes \ac{musa} an efficient \ac{gf} \ac{s-noma} scheme \cite{yang2017uplink}.
%. Additionally, the
%possibility of collision in \ac{gf} \ac{musa}
%is small since a large number of
%spreading sequences can be accommodated

\subsubsection{\ac{noca}}
employs the non-\ac{lds} \ac{lte}-defined low cross-correlation sequences for symbol-level spreading.
Frequency-domain and/or time-domain spreading can be achieved through resource mapping. The former leads to \ac{mc}-\ac{cdma}, and the latter results in \ac{mc}-direct sequence-\ac{cdma}.
\ac{noca} can be developed with small changes in the \ac{lte} standard. Also, it is
effective to establish a
quasi-synchronous channel such that the \ac{bler} is almost the same as for the synchronous case, when the uplink synchronization is kept within one  cyclic prefix \cite{yang2017uplink}.

{\it{3) {\ac{ncma}}}}: {
Grassmannian sequences are employed as user specific
sequences in \ac{ncma}, and aim
at maximizing the minimum chordal distance between
codeword pairs. Grassmannian sequences
represents one
kind of Welch-bound equality sequences; hence, they are the optimal sequences
maximizing the ergodic sum capacity
for equal received power. However, the elements of the Grassmannian sequences are irregular complex values,
which may increase the complexity for hardware implementation \cite{yang2017uplink}.

{\it{4) {\ac{goca}}}}
 is considered as a long sequence-based \ac{s-noma} scheme, which exploits a dual spreading sequence. The spreading sequences in \ac{goca} have a two-stage
structure, which enables group separation through a set of non-orthogonal sequences and device separation within a group through
a set of orthogonal sequences.
\ac{goca} is an enhanced version of \ac{rdma}.
The main
difference between \ac{goca} and \ac{rdma} is that the
former employs orthogonal sequences to spread
the modulation symbols into shared time and frequency
resources after repetition coding in each group.
By employing the \ac{sic} receiver, \ac{goca} can support a high \ac{ovf} due to the large \ac{dof} of the non-orthogonal sequences and
a significant \ac{mui} reduction due to the orthogonal sequences in the same group \cite{3gp4}.

\subsection{Delay Pattern \ac{s-noma}}
The main idea behind the delay pattern \ac{s-noma} is to separate devices in the delay domain.
%Intentional transmission delay provides another dimension for device separation.
%Hence, the deliberate transmission delay is used as device-specific signature.
In this scheme, devices with identical linear spreading sequence are distinguished from their delay at the \ac{ap}.\footnote{Correct decoding is possible as long as two transmission frames do not arrive at the exact same moment.} It should be mentioned that this scheme is usually employed in \ac{gf} \ac{s-noma} and in the \ac{nora} procedure of \ac{rg} \ac{s-noma}.
Fig. \ref{fig:mul776501yty} (solid box) illustrates the concept of delay pattern \ac{s-noma}.

The \ac{rpma} is one of the schemes in this category.
It differs from the \ac{cdma} in two ways: firstly, all devices in a group employ the same Gold code to spread the bits before transmission; secondly, the transmissions are not synchronized and start with a random delay.
%This results in higher \ac{sinr} in \ac{rpma} since the \ac{mui} seen by any device is less due to the random delays.
%While \ac{cdma} is prone to \ac{mui} for the whole duration of the transmission,
%In \ac{rpma}, the random delay is used as device-specific signature, and the good correlation properties of the Gold code allow correct decoding of the transmissions as long as two transmission frames do not arrive at the exact same moment.
%On the other hand, while in \ac{cdma} the uplink time slot
%can be as short as the time it takes to transmit a frame, in
%\ac{rpma} the time slot needs to allow extra time for the time delay of
%each transmission. Thus,
%by extending the time slot, the \ac{sinr} increases at the expense of lowering
%the data rate.
\ac{rpma} supports dynamic \ac{sf} selection without the intervene of the \ac{ap}. A dynamic \ac{sf} is important in order to minimize the power consumption and enable adaptive transmission.
% in terms of bit rate.
For example, when the channel condition is unfavourable, a higher \ac{sf} can be used to reduce the power consumption. On the other hand, when the channel condition permits, a lower \ac{sf} can be used to transmit at higher rates \cite{myers2010random}.

%It is worth noting that devices in \ac{rpma} employ open-loop power control mechanism to change their spreading sequence.

\begin{figure}[h!]
\centering
\hspace{-3em}
\subfloat[MPA receiver.]{%
  \includegraphics[width=9cm]{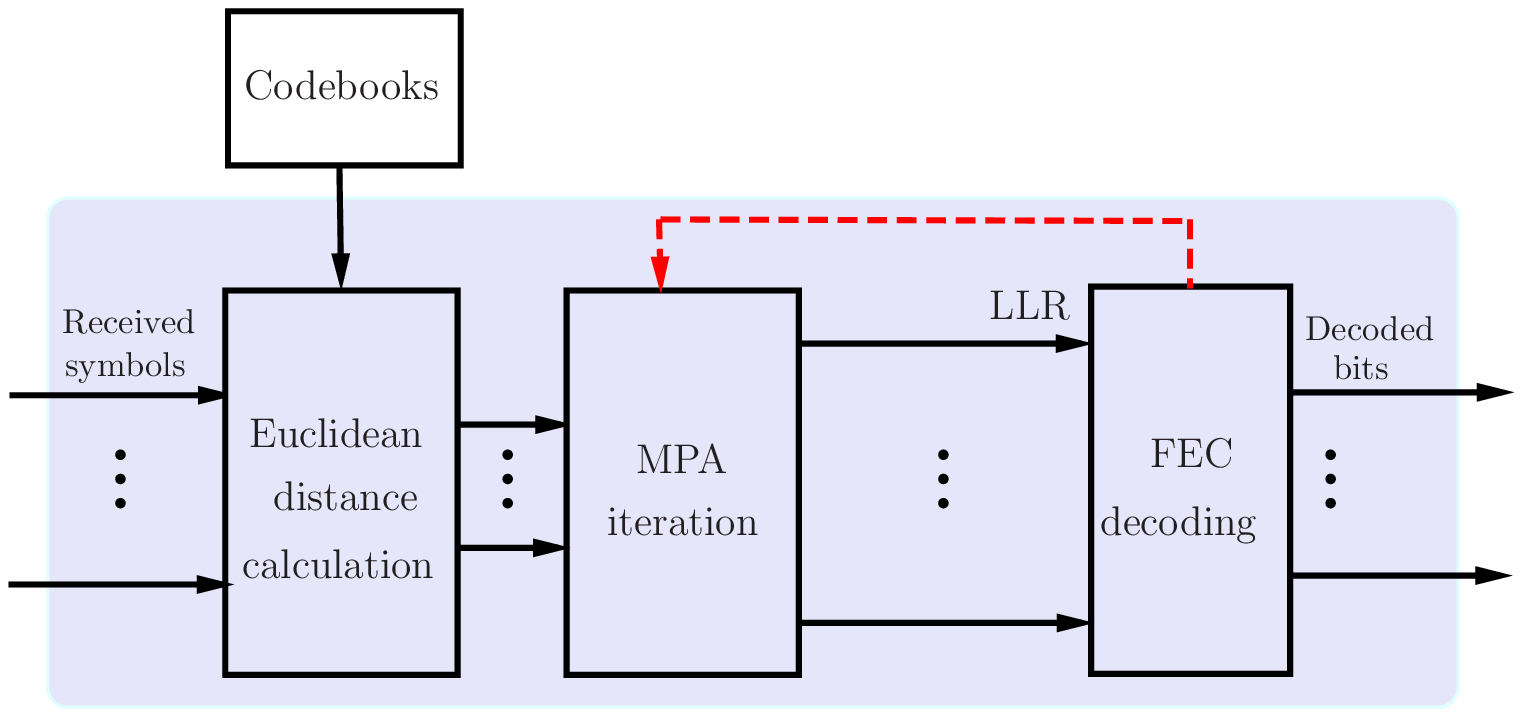}%
  \label{fig:evaluation:revenuewewe}%
  }\
  \vspace{2em}
  \hspace{-2.2em}
\subfloat[SIC (dot blue lines) and PIC (red dash lines) receivers.]{%
  \includegraphics[width=10.3cm]{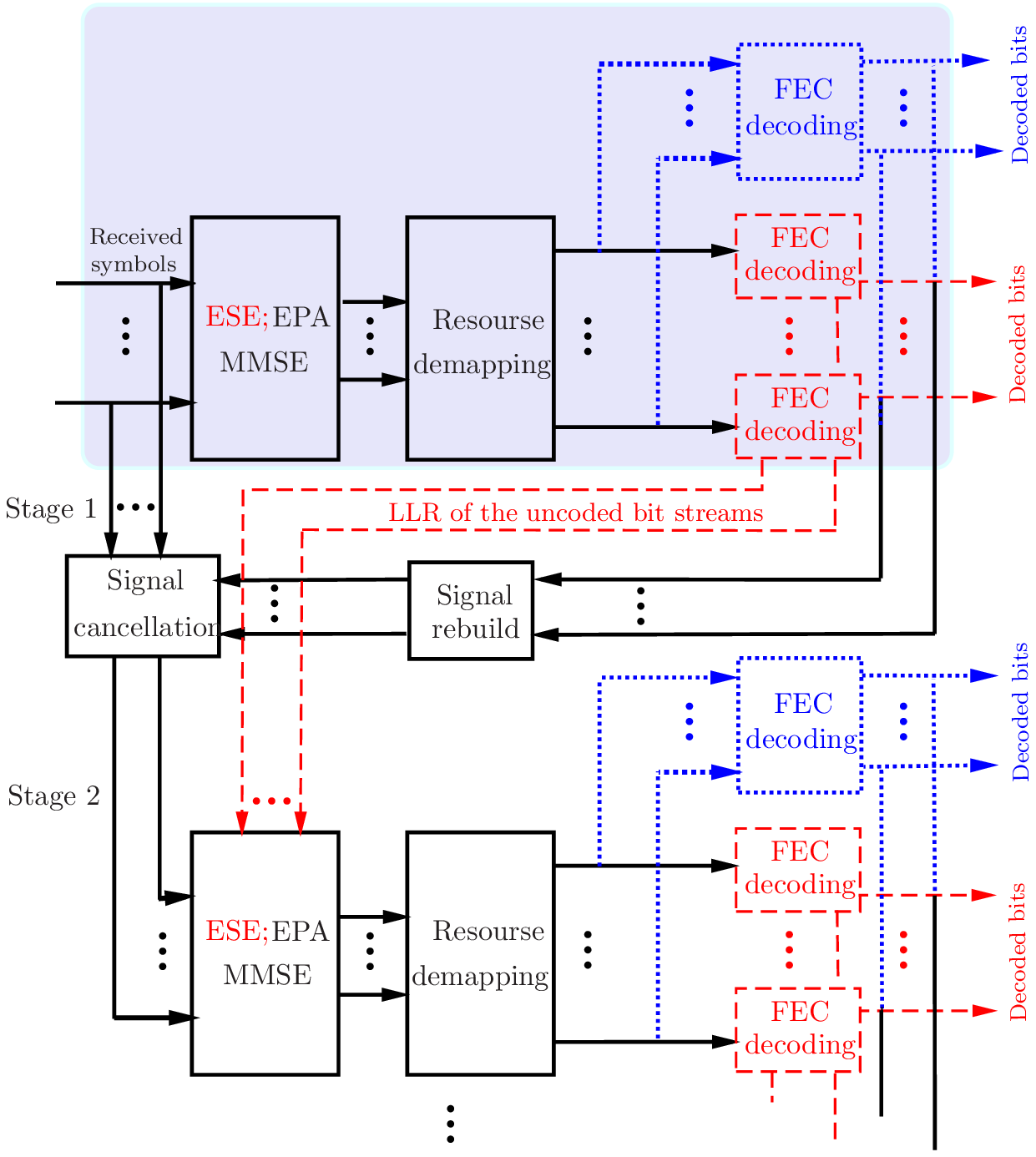}%
  \label{fig:evaluation:revenue}%
}
\caption{Block diagrams of \ac{mud} receivers for \ac{s-noma}.}\label{fig:mul776501op}
\end{figure}

\subsection{Scrambling Pattern \ac{s-noma}}
The main characteristic of the scrambling pattern \ac{s-noma} is that large \ac{dof} for device separation can be obtained through
the combination of very low-rate \ac{fec} and device-specific scrambling sequences.
The low-rate \ac{fec} codes result in high coding gain, and the long low cross-correlation scrambling sequences reduce the \ac{mui}.
Similar to interleaving pattern \ac{s-noma}, the inverse of the code rate represents the \ac{sf}.
A scheme in this category is \ac{rsma}.

{\it {\ac{rsma}}}
employs the combination of very low-rate \ac{fec} and long scrambling sequences
designed for uplink wideband \ac{cdma} (and optionally different interleavers) to spread a group of different devices' signals over
the entire orthogonal resources assigned to the group. Spreading of bits to the entire resources results in high diversity order, which enables decoding at negative \ac{sinr} \cite{yang2017uplink}.

By changing the resource mapping configuration, \ac{sc}-\ac{rsma} and \ac{mc}-\ac{rsma} can be achieved.
The former is optimized for low-power consumption and coverage extension
for short packet transmission. Additionally, it employs low \ac{papr} modulations and enables asynchronous access.
On the other hand, the latter is optimized for low-latency access  \cite{yang2017uplink}.
%It should be mentioned that \ac{rsma} can be employed  with either \ac{gf} or \ac{rg} \cite{yang2017uplink}.

\section{MUD in \ac{s-noma}}
The employed receivers in the uplink \ac{s-noma} are categorized as:
\ac{mpa}, \ac{epa} \ac{sic}, \ac{mpa}-\ac{sic}, \ac{pic}, and \ac{pn} array despreader \cite{3gp1,myers2010random}.
%Fig. 4 illustrates the block diagrams of such receivers, and their description is provided in the following.

\subsection{\ac{mpa} Receiver}
The optimum receiver for \ac{s-noma} is the joint \ac{map} receiver, which gives zero probability of error in the absence of noise and does not suffer from the near-far problem.
However, it exhibits high computational complexity.
%, as it has shown to be an NP-complete problem.
In order to achieve a performance close to that of the \ac{map} receiver with significantly lower computational complexity, the iterative \ac{mpa} receiver can be employed (Fig. \ref{fig:evaluation:revenuewewe}) for
the \ac{s-noma} schemes with \ac{lds} signatures, such as \ac{scma} and \ac{pdma}.
This \ac{mud} receiver jointly decodes all devices by employing graphical models, which
capture the kinds of independence
and factorization structure due to the sparse structure of the signatures to reduce the complexity of the joint \ac{map} receiver.

The decoding in the \ac{mpa} receiver is performed on the Tanner graph constructed by the codebook, where a \ac{fn} represents an orthogonal resource and a \ac{vn} represents the symbol of a device.
The inputs to the \ac{map} receivers are: the received signal on each orthogonal resource, channel estimate on each resource from each device, and noise power estimate on each resource.
The decoding procedure starts with the initial conditional probability calculation at each \ac{fn}. Then, it enters message passing iterations between FNs and VNs along the edges.
There are two steps for each iteration, referred to as \ac{fn} update and \ac{vn} update, which are done independently for each FN-VN pair.
After enough iterations, the probability guess of each codeword is obtained at a \ac{vn} and then is employed for \ac{llr} calculation.
Finally, the achieved \ac{llr}s are directly input to the turbo decoder.
%After a certain number of iterations, the \ac{llr} for coded bits is calculated and given as input to the turbo decoder (or any other
%\ac{fec} decoder) directly.
%after the MPA iteration.
The \ac{mpa} receiver requires no constraint on the \ac{sinr} difference among devices at the receiver, and is robust to the channel correlation between devices. However, its complexity is higher than that of \ac{sic} and \ac{pic} receivers \cite{3gp1}.

\subsection{\ac{epa} Receiver}
This receiver is a graphical-based \ac{mud} algorithm. Compared to the \ac{map} receiver, the \ac{epa} receiver
approximates
the sophisticated posterior distributions in the message update steps with simple distributions
through  distribution  projection by using the Kullback-Leibler divergence. The \ac{epa} receiver is used in \ac{s-noma} schemes with sparse signature, such as \ac{scma}, to archive near \ac{map} performance with lower computational complexity.
By employing the \ac{epa} receiver, the  complexity order
of
\ac{scma} decoding
is  reduced  to
linear complexity,
i.e., it
only scales linearly with
the codebook size
and the  average  degree  of  the  \ac{fn}s
in the factor graph
representation \cite{meng2017low}.

\subsection{\ac{sic} Receiver}
This is the simplest, lowest complexity receiver for \ac{s-noma}, and is particularly appropriate for schemes based on spreading sequence and scrambling pattern, such as \ac{musa}, \ac{noca}, \ac{ncma}, \ac{goca}, and \ac{rsma}.
%In general, short spreading sequences that have low cross-correlation facilitates \ac{sic}
%at the receiver.
The \ac{sic} receiver
decodes devices one after another by treating all the other undecoded devices as interference when decoding one target device.
%This involves serially canceling the interfering devices from the outputs of the \ac{mf}s in order of decreasing power.
At each stage, one or multiple devices with highest \ac{sinr} are decoded. Then, the decoded signals are reconstructed and cancelled out in the next stage (dot blue lines in Fig. \eqref{fig:evaluation:revenue}).
The SIC receiver exhibits low \ac{bler} when the \ac{sinr} levels of devices are significantly different. However, its performance degrades when this \ac{sinr} difference is trivial, because of the error propagation.
Hence, power back-off mechanisms are required to provide different received powers at the \ac{bs}. This is a major challenge in \ac{gf} \ac{s-noma} schemes.
In order to improve the performance of the \ac{sic} receiver, linear receivers, such as \ac{mmse} are employed during each stage of the \ac{sic} receiver \cite{3gp1}. Recently, \ac{epa}-\ac{sic} has also been developed.

 \subsection{\ac{mpa}-\ac{sic} Receiver}
 The combination of the \ac{mpa} and \ac{sic} receivers benefits of their advantages in terms of near-optimal performance and low-complexity, respectively, and is employed  in \ac{s-noma} with \ac{lds} signatures, such as \ac{scma} and \ac{pdma}.

 In such a receiver, the \ac{mpa} is first applied to a limited number of devices in order of decreasing power, so that the number of colliding devices over each orthogonal resources does not exceed a given threshold value, $K_f$. Then, the successfully decoded devices by the \ac{mpa} are removed by \ac{sic} and the procedure continues until all devices are successfully decoded or no device gets successfully decoded by \ac{mpa}. Since MPA is used for a limited number of devices instead of all devices in each stage, the decoding complexity is significantly reduced.
It is worth noting that the \ac{sic}-\ac{mpa} receiver becomes \ac{mpa} when $K_f$ is equal to the number of active devices, and it reduces to the \ac{sic} receiver when $K_f=1$ \cite{3gp1}.
\begin{table*}
\centering
  \caption{Features of \ac{s-noma} schemes.}\label{pdmarr}
\begin{tabular}{c}
  \hspace{1cm}
 \includegraphics[width=17cm]{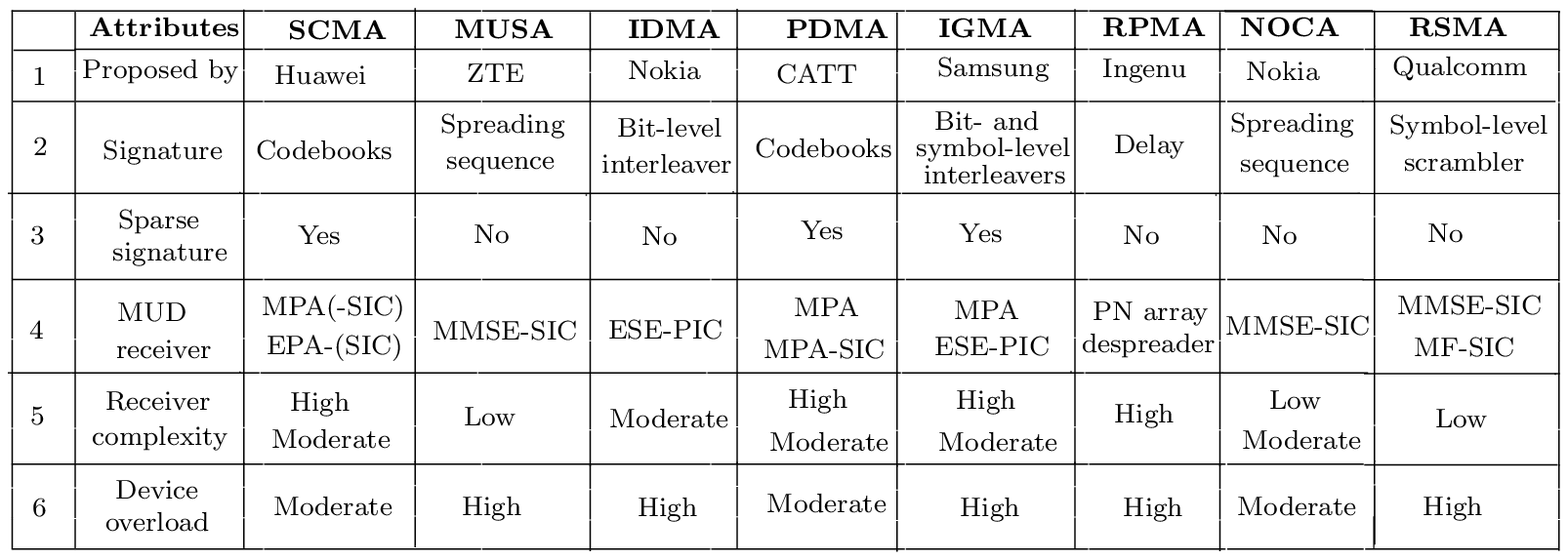}
\end{tabular}
\end{table*}

\subsection{\ac{pic} Receiver}
Opposite to the SIC receiver, the PIC receiver decodes devices in parallel.
At each stage, all devices are decoded and the results are fed back as prior information to the next stage (red dash lines
in Fig. \eqref{fig:evaluation:revenue}).
In general, \ac{pic} exhibits less delay when compared with \ac{sic}, while its complexity is higher.
Linear \ac{mud} receivers, such as \ac{mmse} and \ac{ese} can be employed at each stage of the \ac{pic} receiver to improve the performance \cite{3gp1}.

ESE-PIC is a soft-cancellation-based receiver, mostly used by \ac{s-noma} schemes with bit-level operations like \ac{idma} and \ac{igma}, to provide robust performance for high \ac{ovf}.  It first detects transmitted symbols via \ac{ese} detection;
then, the detected symbols are
deinterleaved and decoded in parallel to acquire coding gain. The output information from the decoder is finally sent
back to the \ac{ese} to aid symbol detection \cite{ping2006interleave}.

\subsection{PN Array Despreader}
This receiver is employed for \ac{s-noma} schemes based on delay patterns, such as \ac{rpma}.
Since the \ac{ap}
has no knowledge of the intentional delays that
the devices add to the beginning of the frame in \ac{gf} \ac{s-noma},
a PN array despreader is employed to despread all the received waveforms through brute-force operation for all possible
chip arrival times and \ac{sf}s.
Each
chip hypothesis is despread, deinterleaved,
decoded, and then checked
via a \ac{crc} to determine the valid frames \cite{myers2010random}.

\vspace{2em}
{\it{Summary}:}
Table \ref{pdmarr} summarizes the features of the discussed \ac{s-noma} schemes. In particular,
the spreading signature, \ac{mud} receiver, complexity of receiver, and \ac{ovf} are provided.
\begin{figure}
\centering
\includegraphics[width=9cm]{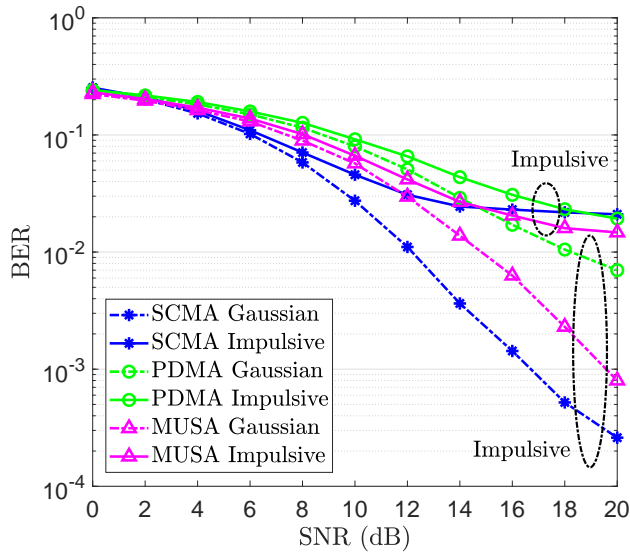}%
\caption{Performance degradation of the \ac{s-noma} schemes in the presence of impulsive noise.}\label{fig:mul776501EWWWWWQ}
\end{figure}

\section{Performance of the \ac{s-noma} Schemes in Impulsive Noise}
In \ac{s-noma},
when the received superimposed signal is impaired by impulsive noise in time domain or by a narrow-band interferer/tone-jamming in frequency domain, some of the time domain chips or some of the subcarriers in the frequency domain may be completely corrupted, which results in performance degradation.

In this section, we evaluate the performance of the \ac{s-noma} schemes
 in terms of \ac{bler}.
We characterize the impulsive noise by employing the Bernoulli-Gaussian
model \cite{pighi2009fundamental}, with the probability density
function $f_{\rv{w}}(x)=(1-\epsilon)f_{{\rv{g}}_1}(x)+\epsilon f_{{\rv{g}}_2}(x)$, where $0 < \epsilon <1$, and $f_{{\rv{g}}_1}(x)$ and $f_{{\rv{g}}_2}(x)$ are zero-mean
complex Gaussian distributions with variances $\sigma_{{\rv{g}}_1}^2$
and $\sigma_{{\rv{g}}_2}^2$.
We set $\epsilon=0.01$ and $\sfrac{\sigma_{{\rv{g}}_2}^2}{\sigma_{{\rv{g}}_1}^2}=100$.

Fig. \ref{fig:mul776501EWWWWWQ} illustrates the \ac{bler} of the \ac{scma}, \ac{pdma}, and \ac{musa}, employing quadrature phase-shift-keying modulation in flat Rayleigh fading, while being contaminated by Gaussian and impulsive noise at the receiver.
The \ac{ovf} is considered $150\%$, while the number of devices and orthogonal resources are 6 and 4, respectively.
The \ac{scma} codebooks  and \ac{pdma} patterns in \cite{wang2015comparison} are employed.
The real and imaginary parts of the complex spreading sequences in \ac{musa} are uniformly drawn from $\{-1,0,1\}$.
As seen, the \ac{bler} of the \ac{s-noma} schemes degrade in the presence of impulsive noise when compared with Gaussian noise. Moreover, the performance degradation in \ac{scma} and \ac{pdma} is higher than in \ac{musa} since the \ac{mpa} receiver is more sensitive to non-Gaussianity
 than the \ac{mmse}-\ac{sic} receiver.

\section{Challenges and Future Trends}
There are several issues that need to be addressed in uplink \ac{s-noma}. Some of these challenges are as follows:

{\it{Joint signature and \ac{mud} receiver design in non-Gaussian noise:}}
The performance of the \ac{s-noma} schemes in their current forms degrades in the presence of non-Gaussian/impulsive noise.
Hence, design of new signatures and \ac{mud} receiver are required to address this problem.

{\it{Activity detection error:}}
The \ac{bler} performance of the \ac{mud} receivers degrades in the presence of the signature identification error in \ac{gf} \ac{s-noma}.
It is important to do a quantitative study of this degradation, e.g., due to the false alarm rate. To the best of the authors' knowledge, such analysis is not available in the literature.

%{\it{Combination of \ac{cs}-\ac{mud} and \ac{s-noma}}:}
%The combination of \ac{cs}-\ac{mud}
%can provide an efficient \ac{gf} \ac{s-noma} that benefits from the joint signature identification, activity detection, and data detection.  Moreover, it enables
%asynchronous and sporadic transmission,
%where active devices, delays, and \ac{csi}
%are unknown.

%{\it{Power consumption}:}
%In uplink \ac{s-noma}, \ac{papr} is a key design parameter that needs to be considered in order to increase the transmission efficiency and reduce the transmission power. Thus,
%\ac{s-noma} schemes with lower \ac{papr} are preferred
%in practical implementations; they also lead to longer battery life time.

{\it{\ac{ovf} in \ac{nora}}:}
The performance of the \ac{rg} \ac{s-noma} scheme in \cite{liang2017non} may be improved by employing \ac{rpma} in the \ac{ra} process.
Opposite to \ac{rpma}, the transmission delay in \cite{liang2017non} is not intentionally added by devices; rather, it is due to the propagation delay. Hence, the \ac{dof} for device separation are restricted by the positions of the devices in the \ac{iot} network. Using \ac{rpma} can provide additional \ac{dof}.

{\it{Channel estimation}:}
The performance of the \ac{mud} algorithms in the \ac{s-noma} schemes heavily relies on the efficiency of channel estimation. New channel estimation methods for \ac{s-noma} are required to be developed.

% is one of the key obstacles in realizing  uplink \ac{s-noma}.
% In general, two \ac{csi} imperfections can be considered for uplink:
%i) channel
%estimation errors, and 2) partial \ac{csi}.
%Channel estimation errors are caused by the imperfect design of channel estimation
%algorithms, which results in device ordering ambiguities in the \ac{sic} and \ac{pic} receivers.
%The impact of channel estimation errors on different \ac{s-noma} schemes needs to be studied, and efficient channel estimators developed.
%The use of partial \ac{csi} is motivated by the fact that small scale multipath fading varies
%much faster than large scale path loss. This means that learning the path loss information
%of the device channels requires less overhead at the transmitter than estimating both multipath
%fading and path loss. The performance degradation of the \ac{s-noma} schemes when only partial \ac{csi} is available at the \ac{ap} needs to be explored.

{\it{\ac{mud} receiver complexity}:}
For the \ac{mpa} receiver, the complexity may
still be high for massive connectivity.
Therefore, new \ac{mud} receivers should be designed to reduce complexity. One approach is using the Gaussian approximation for interference-plus-noise;
this approximation
tends to be more accurate as the connectivity
becomes larger in \ac{iot}.

%{\it{Error propagation suppression}:}
%For a \ac{sic} receiver, the error propagation may
%degrade the performance of some devices. Thus,
%at each stage of \ac{sic}, nonlinear detection algorithms
%with higher detection accuracy can be considered
%to suppress the error propagation.
%This provides a trade-off between complexity and performance.

%{\it{Higher-layer designs}:}
%Higher-layer considerations play an
%important role to decrease the control signaling in \ac{rg} \ac{s-noma}; examples are
%the introduction of connectionless one-shot
%transmission modes to enable longer sleep cycles, and
%traffic prediction approaches to efficiently support
%quasi-periodic traffic.

\section{conclusion}
%Massive connectivity is a major concern in IoT applications, and \ac{noma}
%is a potential mechanism to support a large number of devices with a limited number of
%resources.
Uplink \ac{s-noma} schemes for massive \ac{iot} were reviewed in this article.
%  with focus on
%their capability to provide massive connectivity.
The modifications of \ac{gf} and
\ac{rg} schemes were introduced for \ac{s-noma}. For different \ac{s-noma} schemes, the
design of signatures was presented, along with the transmitter and receiver structures, and receiver
complexity and performance.
% For example, \ac{scma} provides a high
%\ac{ovf} with a complex \ac{mud} algorithm at the receiver, while
%\ac{musa} and  \ac{rsma} support large overloading
%factors and have relatively reduced complexity receivers.
Moreover, the performance of the \ac{scma}, \ac{pdma}, and \ac{musa} schemes in the presence of impulsive noise was studied through simulation, and it was shown that they are generally sensitive to impulsive noise. Finally, various
directions for future investigation were provided.

\bibliographystyle{IEEEtran}


\end{document}